\documentclass[10pt,a4paper,twoside]{article}

	\usepackage{amssymb}
	\usepackage{amsmath}
	\usepackage{bm}
	\usepackage{graphicx}
	\usepackage{subfigure}
	\usepackage{hyperref}
	\usepackage{color}
	\usepackage[font=small,labelfont=bf]{caption}

	\usepackage{booktabs}
	\usepackage{array}
	\usepackage{enumitem}
	\usepackage{cite}
	\usepackage[affil-it,auth-lg]{authblk}
	\usepackage{hyperref}
	\usepackage{eso-pic}

\begin{document}

\title{Binary black hole growth by gas accretion in stellar clusters}

\author[1,2\footnote{Visiting scientific collaborator, April-May 2018}]{Zacharias Roupas}
\affil[1]{Department of Mathematics, University of the Aegean, Karlovassi 83200, Samos, Greece} 

\author[2]{Demosthenes Kazanas}
\affil[2]{Astrophysics Science Division, NASA Goddard Space Flight Center, Greenbelt, MD 20771, USA} 
	
\date{\vspace{-5ex}}

\maketitle

\begin{abstract}
        We show that binaries of stellar-mass black holes formed inside a young protoglobular cluster, can grow rapidly inside the cluster's core by accretion of the intracluster gas, before the gas may be depleted from the core. A black hole with mass of the order of eight solar masses can grow to values of the order of thirty five solar masses in accordance with recent gravitational waves signals observed by LIGO.
        Due to the black hole mass increase, a binary may also harden. 
        The growth of binary black holes in a dense protoglobular cluster through mass accretion indicates a potentially important formation and hardening channel.
\end{abstract}

\hspace{10pt}


\section{Introduction}

\indent 
The majority of gravitational wave (GW) signals produced by binary black holes (BBH) and observed by LIGO \cite{Abbott_2016PhRvL.116f1102A, Abbott_2016PhRvL.116x1103A,Abbott_2017PhRvL.118v1101A,Abbott_2017ApJ...851L..35A,Abbott_2017PhRvL.119n1101A,LIGO_2018arXiv181112907T} involve high stellar mass black holes (BHs) 
$\sim (20-50)M_\odot$.
In particular, recently LIGO \cite{LIGO_2018arXiv181112907T} released new data indicating that $75\%$ of the BHs, observed in $10$ BBH GW signals, have masses greater than $20M_\odot$.
We investigate here a possible origin of such high mass BHs.
In the, so called, dynamical channel \cite{Abbott_2016ApJ...818L..22A}, the BBH is assumed to be formed and get hardened in the dense environment of a stellar cluster, such as a globular cluster (GC) or nuclear star cluster (NSC), by dynamical processes. We examine the possibility that a BBH can grow by accretion of primordial gas during the cluster's early formation era.

Globular clusters seem to contain little or no interstellar gas today. However, there is accumulated evidence that they underwent prolonged star formation early in their lifetimes \cite{Gratton_2012A&ARv..20...50G} and therefore should have started life as a dense gas cloud, called a protoglobular cluster or primordial cluster. It is uncertain to what extent these primordial clusters share the same properties as the young massive clusters (YMC) observed today (see \cite{Portegies_2010ARA&A..48..431P} and references therein). For a vast range of values for the stellar formation efficiency $\sim (10-90)\%$, the mass of primordial gas remaining after the first formation event in a primordial cluster is of the same order of the cluster's stellar mass. Therefore, there is a huge gas reservoir, available for accretion, immediately after the first formation event. The BBH accretion efficiency will depend strongly on the relevant timescales of the depleting and accreting processes. 

\indent In star-forming regions, the phenomena of radiation, stellar winds and energetic explosions are expected to clear away the surrounding gas within a few $Myr$ from the onset of formation (e.g., \cite{Voss_2010A&A...520A..51V,Galvan-Madrid_2013ApJ...779..121G,Krumholz_2014prpl.conf..243K}).
However, this is not necessarily true for sufficiently massive clouds. Since feedback processes are proportional to the total mass, while the gravitational binding energy to the square of the mass, there should exist a critical mass, depending on the size of the cloud, above which the feedback processes become ineffective. Sufficiently compact clusters such as NSCs and massive GCs are candidate regions in which  this may occur. In fact, it has been shown that a key parameter controlling gas expulsion is indeed compactness, namely the mass over size of the cluster \cite{Krause_2012A&A...546L...5K,Krause_2016A&A...587A..53K,Silich_2017MNRAS.465.1375S,Silich_2018MNRAS.478.5112S}.  

 The precise mechanism for gas depletion in primordial GCs is still unknown, 
 while proposals abound in the literature with most of them focusing on stellar winds and supernovae explosions \cite{Spergel_1991Natur.352..221S,Thoul_2000LIACo..35..567T,Fender_2005MNRAS.360.1085F,Moore_2011ApJ...728...81M,Herwig_2012ApJ...757..132H,Krumholz_2014prpl.conf..243K,Krause_2016A&A...587A..53K,Silich_2017MNRAS.465.1375S,Silich_2018MNRAS.478.5112S}. The gas expulsion may dramatically affect the evolution of the cluster \cite{Marks_2008MNRAS.386.2047M,Kruijssen_2012MNRAS.426.3008K} and especially in connection with the presence of multiple stellar populations \cite{DErcole_2008MNRAS.391..825D,Conroy_2012ApJ...758...21C,Renzini_2015MNRAS.454.4197R}. It has been argued that the superbubbles formed by winds and supernova explosions \cite{Bagetakos_2011AJ....141...23B,Krause_2013A&A...550A..49K,Jaskot_2011ApJ...729...28J,Fierlinger_2016MNRAS.456..710F,Yadav_2017MNRAS.465.1720Y}, which are the primary candidates supposed to expel the primordial gas, undergo a Rayleigh-Taylor instability in sufficiently massive ($\gtrsim 10^7M_\odot$) proto-clusters preventing this gas expulsion \cite{Krause_2012A&A...546L...5K}. 
Krause et al. \cite{Krause_2012A&A...546L...5K} propose further that, in this case, the power released by accretion of primordial gas onto dark remnants could be sufficient to expel the gas. Leigh et al. \cite{Leigh_2013MNRAS.429.2997L} have further elaborated the idea and proposed that in any cluster that is able to form massive stars, the primordial gas is depleted exactly due to the accretion onto BHs. They find that accreting BHs can deplete the whole gas reservoir within only as few as $10Myr$.

Here, we do not focus on the effect of the accreting BHs to the gas reservoir, but on the effect of accretion to the BBHs of the cluster. We assume that the gas depletion occurs rapidly in a timescale of few $Myrs$, independently of any specific depletion scenario. We further assume that the cluster can generate at least one stellar mass BBH, that may be found inside the clusters' core $\lesssim 1pc$ within a few $Myrs$ from the onset of the first formation event. This assumption is justified, since most massive stars are formed as binaries \cite{Sana_2012Sci...337..444S} and furthermore they sink rapidly, in relevant timescales, to the center due to mass segregation and dynamical friction \cite{Spitzer_1987degc,B&T_2008gady.book}. In fact, it has been calculated recently \cite{2014MNRAS.441..919L} that accretion of gas (gas damping) accelerates mass segregation resulting to timescale of less than $1Myr$.

Moreover, the timescales considered are much shorter than the $Gyr$ timescale that the recoil mechanism of BBH-BH three-body encounters operate and which may eject BBHs out of the cluster \cite{Sigurdsson_1993Natur}. In addition, following numerous pieces observational evidence of X-ray emitting black holes \cite{2007Natur.445..183M,2011ApJ...734...79B,2010ApJ...721..323S,2011MNRAS.410.1655M,2012Natur.490...71S}, it has become evident
in the last decade that the recoil mechanism is not effective in dense GCs, which seem to contain a significant population ($\lesssim 1000$) of BHs and BBHs in their core (see \cite{Morscher_2015ApJ...800....9M} and references therein). 

Our main result below is that in a protoglobular cluster a sufficient amount of primordial gas can be accreted on a BBH,  before the gas may be depleted, to increase the mass of each BBH member to values $\sim 30 M_\odot$, consistent with high mass BBH LIGO detections \cite{LIGO_2018arXiv181112907T}.
We further calculate, in case of isotropic accretion, the degree that a BBH gets harder (with the ``hardness" $\chi(t)$ defined below in equation (\ref{eq:chi})) due to accretion. 
We discuss this issue along with our conclusions in Section 3.

The present analysis suggests that in addition to the
two primary channels \cite{Abbott_2016ApJ...818L..22A} for formation and hardening of BBH that may be detected by LIGO (the ``dynamical'' 
in dense stellar environments 
and the ``isolated'' channel  
in isolated environment),
an additional formation and hardening channel of BBH operates due to accretion of gas in dense primordial stellar clusters.

In the next section we present our results, while the basic calculations are given in the appendices. In Appendix \ref{sec:acc} we briefly describe the mechanism of isotropic accretion, in Appendix \ref{sec:hard} we show our calculation of the hardening of a BBH due to isotropic accretion and in Appendix \ref{sec:evolution} we present the equations that describe the evolution of the BBH members. In Appendix \ref{sec:ion} we discuss the timescale of ionization of a soft binary versus that of the hardening due to accretion.

\begin{figure*}[tbh]
\begin{center}
        \subfigure[$a(0) = a_h(0)$ and $m_\bullet(t_f)=30M_\odot$]{ \label{fig:rho_min_t}
        \includegraphics[scale = 0.5]{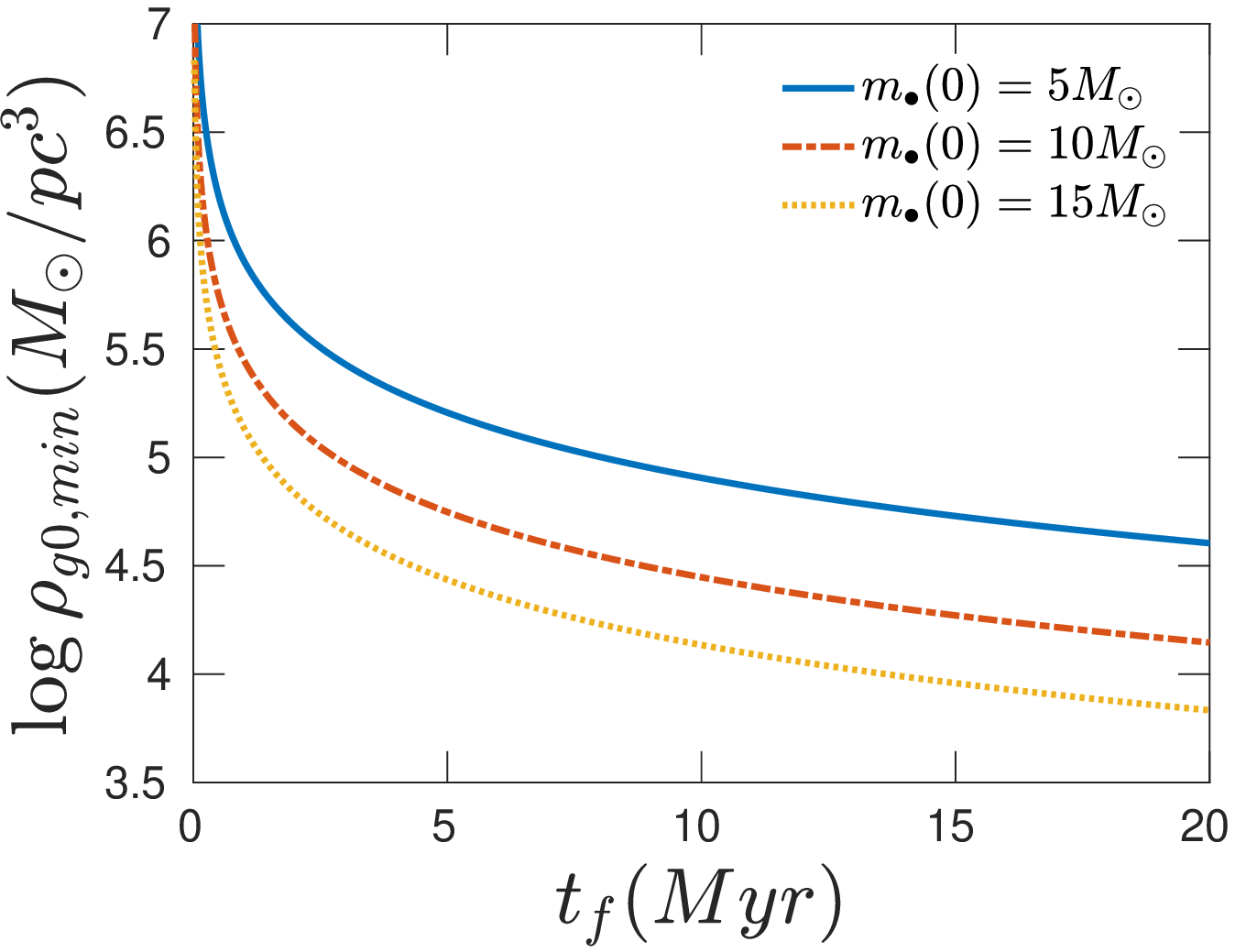}  }
        \subfigure[$t_f = 5Myr$ and $m_\bullet(t_f)=30M_\odot$]{ \label{fig:rho_min_a}
        \includegraphics[scale = 0.5]{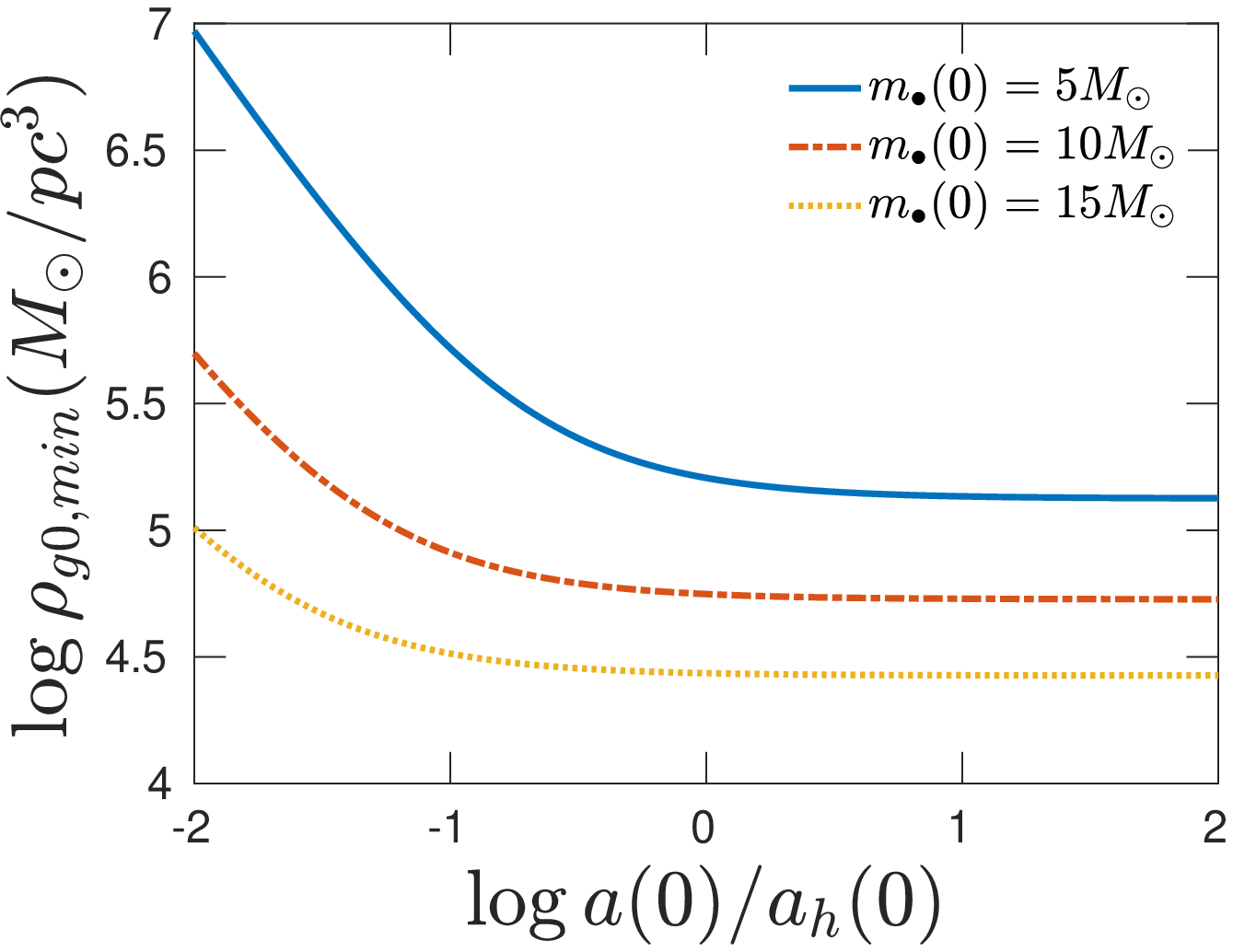}  }
        \caption{
        Minimum initial density of the gas ($\rho_{g0,min}$), given in Equation (\ref{eq:rho_min}), required for a BBH with equal mass BH members so that each member reaches mass $30M_\odot$ by Bondi accretion inside this gas' density environment and until time $t_f$ when the gas is completely depleted, assuming a constant mass loss rate of the gas (\ref{eq:rho_g_t}). For an exponential gas loss law (\ref{eq:rho_g_t-exp}), the time $t_{99}$ needed for $99\%$ depletion is $t_{99} = 2.3 t_f$, for the same values of all other parameters. We also assume in both figures $c_s = 5km/s$, $\sigma = v_c = 6km/$ where we denote, respectively, the speed of sound of the gas, the velocity dispersion of the stars and the velocity of the binary with respect to the cluster. We consider three cases, each member having initial mass $m_0 = 5M_\odot$ (blue solid line), $10M_\odot$ (red dashed-dotted line) or $15M_\odot$ (yellow dashed line). 
        \textit{Left:} We assume an initially slightly hard binary with $\chi(0) = 1 \Leftrightarrow a (0) = a_h(0)$. It is evident that in order for a BH with initial mass $\sim (5-10)M_\odot$ to grow up to $30M_\odot$ within $t_f = 5Myr$, assuming by this time there is complete gas depletion, would require an initial gas' density $\sim (5-15)\cdot 10^4M_\odot/pc^3$. These minimum density values correspond to the exponential gas loss law with $99\%$ depletion at time $t_{99} = 11.5Myr$. The density values are half for $t_f = 10Myr$ for both linear and exponential cases. We took into account the hardening of the binary during accretion. 
        \textit{Right:} We calculate $\rho_{g0,min}$ w.r.t. the initial hardness expressed by the quantity $\log\chi^{-1} (0)$ for $t_f = 5Myr$ in the linear case of gas loss. This time corresponds to $t_{99} = 11.5Myr$ in the case of exponential gas loss. It is evident that soft binaries accrete gas more effectively than very hard ones. A hard binary $a (0)/a_h(0) = 0.1 $ with BH members each of mass $(5-10)M_\odot$ requires $\rho_{g0,min}=(8-50)\cdot 10^4M_\odot/pc^3$ while a soft one $a(0)/a_h(0)=10 $ requires only $(5-13) \cdot 10^4M_\odot/pc^3$. By the time the gas is depleted a soft binary gets hard in case of isotropic accretion as demonstrated below, in Figure \ref{fig:ma_t_GW150914}. 
        \label{fig:rho_min}}
\end{center} 
\end{figure*}

\section{Results}

\begin{figure*}[!tbh]
\begin{center}
        \subfigure[Bondi, hard]{ \label{fig:ma_t_hard_exp_GW150914}
        \includegraphics[scale = 0.5]{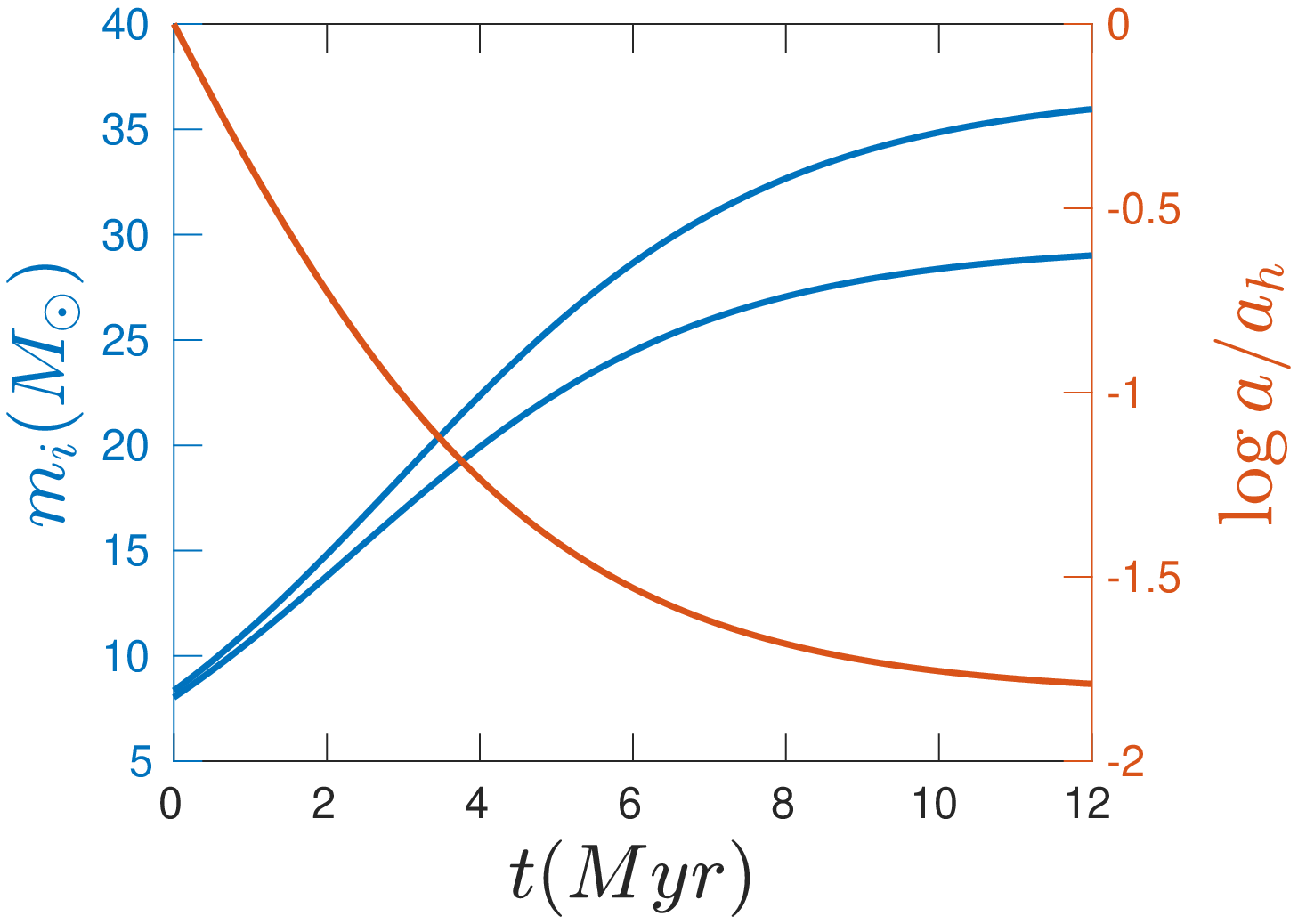}  }
        \subfigure[Eddington, hard]{ \label{fig:ma_t_hard_Edd_GW150914}
        \includegraphics[scale = 0.5]{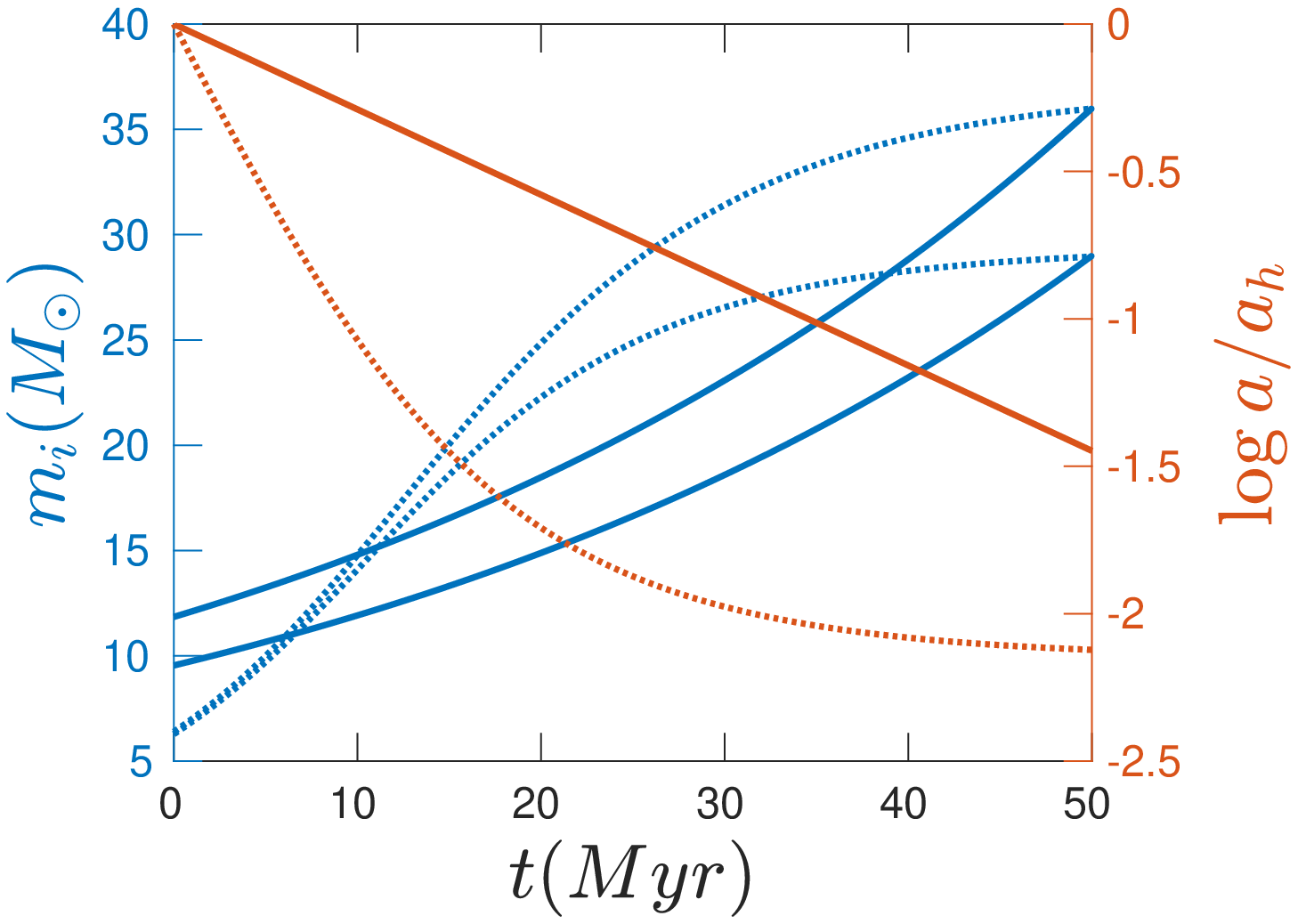}  }
        \\
        \subfigure[Bondi, soft]{ \label{fig:ma_t_soft_exp_GW150914}
        \includegraphics[scale = 0.5]{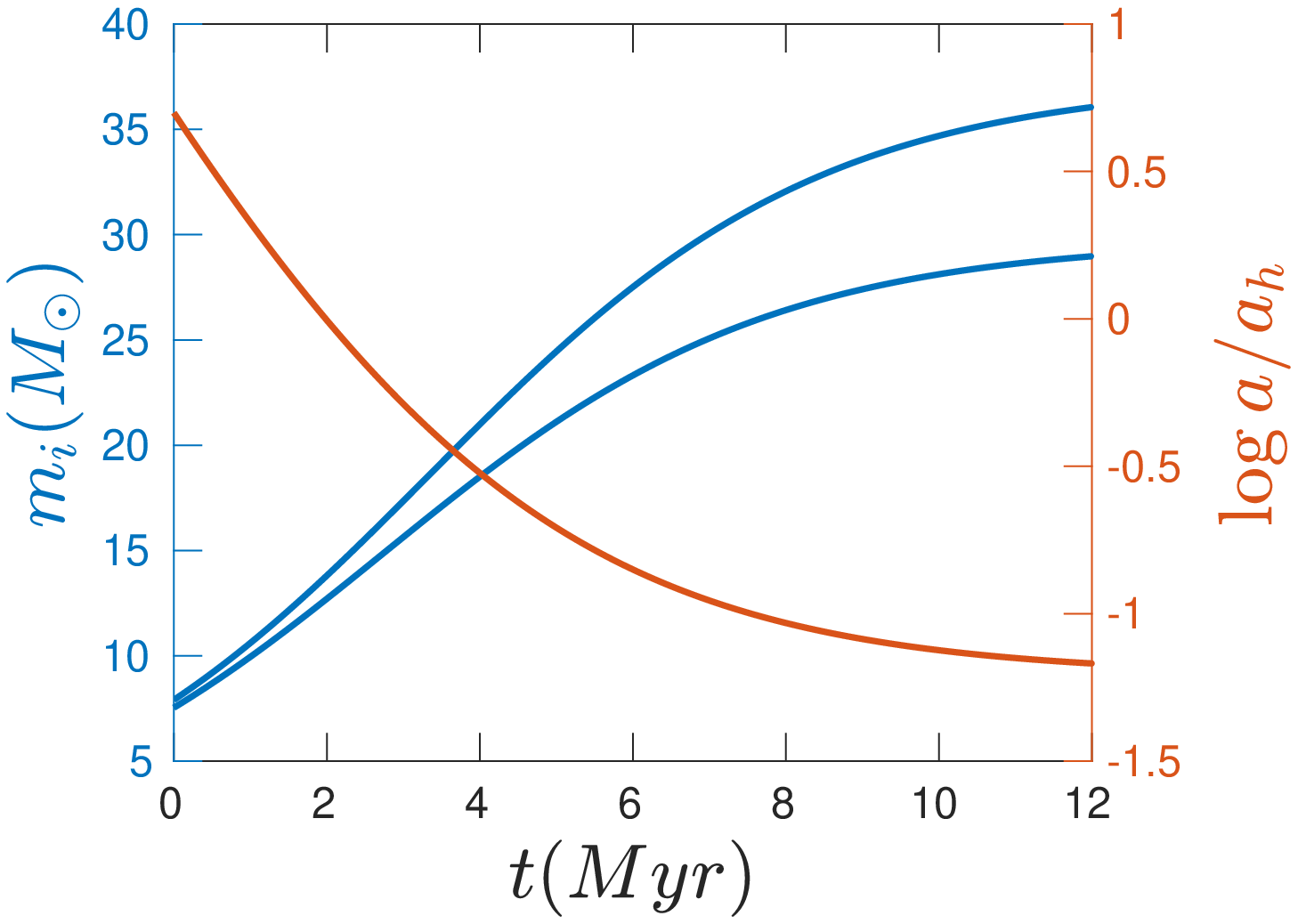}  }
        \subfigure[Eddington, soft]{ \label{fig:ma_t_soft_Edd_GW150914}
        \includegraphics[scale = 0.5]{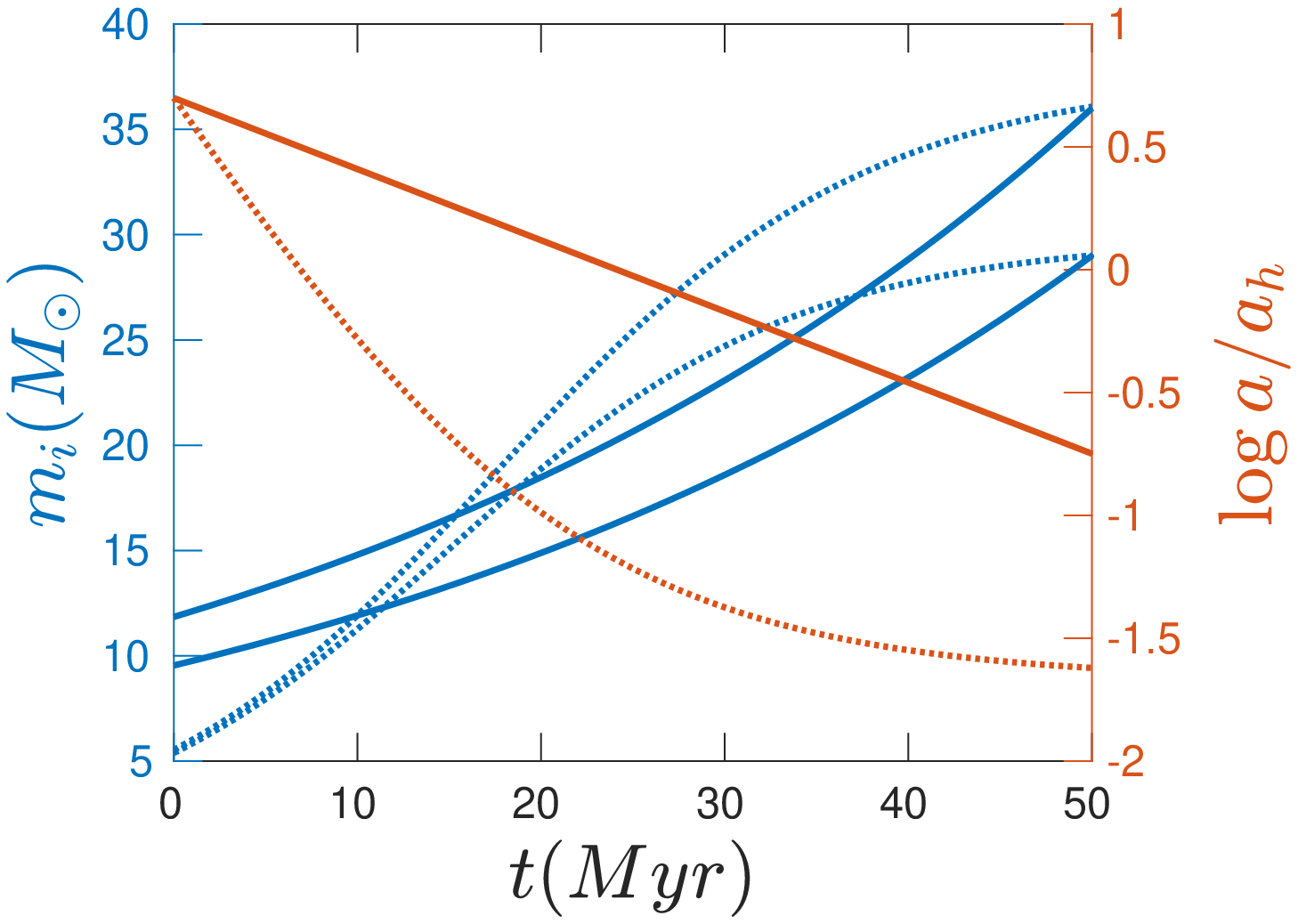}  }
        \caption{Evolution of BH masses $m_1$, $m_2$ (red lines) and BBH hardness (minus logarithm) $\chi(t) = a_h(t)/a(t)$ (blue lines) plotted until the time the gas reservoir of a primordial cluster is completely depleted for an exponential loss law (for a linear law the depletion time for the same mass growth is about half) so that the BHs attain the mass values of the GW signal GW150914, namely $m_1(t_\text{end}) = 36M_\odot$, $m_2(t_\text{end}) = 29M_\odot$. We assume $c_s = 5km/s$, $\sigma = 6km/s$ and the orbital velocity of the BBH w.r.t. the cluster to be constant $v_c =4.5km/s$ which corresponds to a circular orbit of the barycenter at a distance $\sim 0.15pc$ from the center of the cluster. Assuming $v_c = \sigma$ will add about $8Myr$.  
        \textit{Left Panels:} We consider spherically symmetric Bondi accretion and solve the system (\ref{eq:dm1dt})-(\ref{eq:dm2dt}). The initial gas density is $\rho_{g}(0) = 5\cdot 10^4M_\odot/pc^3$. We find that the time needed is $12Myr$ for $a(0)/a_h(0) = 1$, corresponding to the soft-hard boundary, in the upper panel and $m_{1}(0) = 8.35M_\odot$, $m_2(0) = 8.03M_\odot$, while  we consider a soft binary in the lower panel with $a(0)/a_h(0) = 5 \Leftrightarrow \chi(0) = 0.2$ and $m_{1}(0) = 7.91M_\odot$, $m_2(0) = 7.54M_\odot$.
        The hard BBH gets about $62$ harder, while the soft $74$ times harder becoming very hard at the end of the process $\chi(t_\text{end})=14.8$. Both the mass increase and hardening effects are more intense for the soft binary in agreement with Figure \ref{fig:rho_min_a}. 
        \textit{Right Panels:} We consider Eddington-limited accretion in a disk, as in eq. (\ref{eq:Eddington}), depicted with solid lines and we plot in the same graph the Bondi case with dotted lines. We assume initial gas density $\rho_{g}(0) = 10^4M_\odot/pc^3$. The time needed for a similar growth to the Bondi case is about four times larger with respect to to the left panels, namely $50Myr$ with a little higher initial masses $m_{1}(0) = 11.85M_\odot$, $m_2(0) = 9.55M_\odot$. We note that in order to achieve the same final mass in the same time with Bondi accretion, the initial masses are much lower, namely $m_{1}(0) = 6.40M_\odot$, $m_2(0) = 6.29M_\odot$ in the hard case and $m_{1}(0) = 5.54M_\odot$, $m_2(0) = 5.40M_\odot$ in the soft one. In case of angular momentum preservation (or loss to the disk) the BBH gets harder in Eddington-limited accretion as well, though it is less severe than in the Bondi case. The soft binary gets hard attaining $\chi(t_\text{end}) = 5.6$. 
        \label{fig:ma_t_GW150914}}
\end{center} 
\end{figure*}

The isotropic, spherically symmetric accretion is described by the Bondi formula for each BBH member, as in Appendix \ref{sec:acc}. During isotropic accretion, the binary is hardened as we show in Appendix \ref{sec:hard}. The accumulation of mass by BBH members results in a decrease of the separation due to angular momentum preservation.

The formation of a thin disk may, however, significantly decrease the accretion rate. Radiation emission during accretion could negatively impact the accretion rates
due to gas heating and ionization via inverse Compton scattering \cite{King_2003ApJ...596L..27K}. Under this perspective the Bondi accretion analysis corresponds to nearly maximum possible accretion rate. On the other hand, Eddington-limited accretion should be much closer to the true rate if accretion proceeds via angular-momentum redistribution within a disk that radiates \cite{King+Pounds_2003MNRAS.345..657K}. It may be regarded for our problem as corresponding to nearly minimum accretion effectiveness being linearly depending on BH mass as in Eq. (\ref{eq:Eddington}) of Appendix \ref{sec:acc}.

Regarding the gas loss, we investigate two cases; a linear gas loss rate, as in (\ref{eq:rho_g_t}) and an exponential gas loss rate, as in (\ref{eq:rho_g_t-exp}). We do not worry about the cause of the depletion. Our results are independent of any specific depletion scenario as long as its loss rate lies within these two marginal gas loss models.

In Appendix \ref{sec:evolution} we show our calculation of the minimum possible initial gas density $ \rho_{g0,min}$, as in (\ref{eq:rho_min}), (\ref{eq:rho_min-exp}) for hard and soft binaries and for several initial BH masses so that $m_\bullet (t_f) = 30M_\odot$. The result is plotted in Figure \ref{fig:rho_min}. 
We find that $\rho_{g0,min} \sim (10^4-10^5)M_\odot/pc^3$, gas density values relevant to proto-globular clusters. 
The number density of gas in a primordial cluster is estimated to be $\sim 10^6cm^{-3}$ and the temperature a few thousand Kelvin degrees ($\sim 5000K$) \cite{DErcole_2008MNRAS.391..825D,Maccarone_2012MNRAS.423....2M,Calura_2015ApJ...814L..14C}. These values give $c_s\sim 5km/s$ and $\rho_g \sim 5\cdot 10^4M_\odot/pc^3$ for the speed of sound and the gas density, respectively, in agreement with our above result. 
Moreover, the gas density value may be even higher. A primordial cluster is estimated \cite{Fall_2001ApJ...561..751F,Kruijssen_2009ApJ...698L.158K,Conroy_2012ApJ...758...21C} to be $10-100$ times more massive than the resulting GC. Typical galactic GCs have mass $\sim 5\cdot (10^4- 10^5)M_\odot$ (the mean mass of galactic GCs is $\sim 2 \cdot 10^5M_\odot$ \cite{Harris_1998}) and therefore originate from primordial clusters of mass $ 5\cdot (10^5-10^7)M_\odot$. Let us consider a primordial cluster with mass $10^7M_\odot$, and assume that the formation efficiency is $0.4$. If the half mass radius of the gas' distribution is $r_c \sim 3pc$ then the mean gas density inside $r_c$ is $\sim 10^6M_\odot/pc^3$, a value much higher than the one given above. Unfortunately, the exact conditions prevailing in the center of a primordial cluster are unknown.

Then, in Figure \ref{fig:ma_t_GW150914} we present a demonstrative example of a BBH becoming more massive and harder by accretion. In particular, we consider both Bondi and Eddington-limited accretion cases in order to generate the masses of GW signal GW150914, as a representative of the signals involving high ($m_\bullet>20M_\odot$) BH masses. 
These can originate in BBHs with originally light BH members ($\sim 8-10M_\odot$), which increased their masses rapidly, within $\sim 10Myr$ in Bondi accretion and $\sim 50Myr$ in Eddington-limited case, assuming that the gas is $99\%$ depleted by exponential loss within this time.

We note that in case of Bondi accretion a very small initial mass difference between the two BHs grows fast leading to significant final mass difference at the end.  Also, due to dynamical friction that is not taken into account here, the time by which the soft binary becomes hard is overestimated and it should become even harder at the end. 

We find that for the aforementioned parameter values of Figure \ref{fig:ma_t_GW150914}, a soft BBH with $\chi(0) = 0.2$ gets hardened due to mass increase, in case of angular momentum preservation, by a factor  $28-74$ becoming hard, $\chi(t_{\text{end}}) = 5.6-14.8$, at the end of the process, where the lower values correspond to Eddington-limited case and upper ones to Bondi case. Soft binaries that would normally dissolve due to three-body encounters (Heggie-Hills law) they may get hard ($\chi(t_f) > 1$) due to accretion that forces their concentration in the center. 
However, the simultaneous operation of three-body encounters should act counter to this effect, so that not all soft binaries should be subject to such a drastic accretion effect and not under all conditions. 
The timescales of ionization of soft binaries (\ref{eq:ion}) and hardening (\ref{eq:tau_h}) are comparable at least for mildly soft binaries inside the core of a proto-globular cluster with $\rho_\text{gas}$ greater than $\rho_\text{stars}$ (see eq. (\ref{eq:tau_ratio}) in Appendix \ref{sec:ion}).
The inclusion of both dynamical friction and close encounters needs much further investigation and will be presented elsewhere.

\section{Conclusions}

Our basic results are summarized in Figures \ref{fig:rho_min} and \ref{fig:ma_t_GW150914}. We conclude that: 
\begin{enumerate}[noitemsep,nolistsep]
        \item BHs with initial masses $(5-10)M_\odot$, members of a BBH, require initial gas density environment $\sim (10^4-10^5)M_\odot/pc^3$ in order to become as massive as $30M_\odot$ by isotropic accretion of gas within a time of $\sim 12Myr$ for $99\%$ gas depletion following an exponential gas loss. Such gas density values are expected to occur in the centers of primordial clusters, while the aforementioned gas depletion timescales are in direct agreement with simulations of gas depletion due to feedback from star formation in primordial clusters \cite{Calura_2015ApJ...814L..14C} (see Appendix \ref{sec:evolution}). 
        \item 
        The GW signals involving high mass ($\gtrsim 20M_\odot$) BHs
        can originate at BBHs with light BH members, each of mass $\sim 8 M_\odot$, that became sufficiently massive (in particular $36M_\odot$ and $29M_\odot$ for GW150914) by gas accretion in primordial clusters within gas depletion time of $12Myr$ for exponential gas loss and Bondi accretion and an initial gas density $5\cdot 10^4 M_\odot/pc^3$. 
        \item In the case of the more conservative Eddington-limited accretion in a radiative thin disk, the time needed for the BBH to grow at similar proportion to the Bondi case is $\sim 50Myr$ and for slightly higher initial BH masses $\sim 10M_\odot$. The two accretion cases may be regarded as upper and lower limits so that the precise timescale should lie somewhere between $10-50Myr$.
        \item The BBHs get hardened during isotropic accretion due to conservation of angular momentum, proportionally to the third power of mass, as in Equation (\ref{eq:chi_law}).
        \item The mass increase and hardening effects are more intense for soft binaries, which it is possible to become hard due to isotropic gas accretion before being dissolved due to three-body encounters, especially for mildly soft binaries in a stellar or BH subcluster immersed in a much denser gas environment (see Eq. (\ref{eq:tau_ratio}) in Appendix \ref{sec:ion}).
\end{enumerate}

Depending on the conditions prevailing in the primordial cluster and the specifics of the BBH and the environment it is embedded in, there may or may not be formed a gas accretion (mini)disk with some thickness. Recent simulations \cite{2014ApJ...783..134F,2015ApJ...807..131S} show that accretion in the presence of a thin disk is not significantly suppressed compared to the accretion rate expected for a single black hole with the binary mass. Therefore, our results will be valid regarding the mass increase effect even in the case of the presence of a disk, while the precise arithmetic corrections are expected to be small according to 3D simulations \cite{2015ApJ...807..131S}. A qualitative picture may be given as follows. Provided that the disk cannot be stable and grow beyond some size and mass because of (and not only) being immersed inside a dense gas density environment there should operate some form of continuity principle. For any amount of gas mass that enters some prescribed radius which encloses the BBH and the disk, an almost equal amount of mass is accreted onto the BHs instead of accumulating on the disk. Nevertheless, this issue requires further investigation.

Regarding the effect of minidisks to the hardening of the BBH due to accretion, we emphasize that simulations show that orbital angular momentum tends to be removed from the BBH either due to "spiral shocks" \cite{1987A&A...184..173S,2017ApJ...835..199R,2016ApJ...823...81J,2018ApJ...854...84A,2017MNRAS.469.4258T} or magnetohydrodynamic effects \cite{1989ApJ...342..208K,1999AAS...194.9705S}. These effects are small and in any case they accelerate the hardening. According to this literature, our estimation for the hardening rate is in fact a lower limit and it can be even higher in the presence of minidisks. We note, however, that some authors \cite{2017MNRAS.466.1170M} contradict the aforementioned literature, and a general consensus has not as yet been achieved. 

In addition, it has been argued that dynamical friction due to the gas generates torques which tend to harden the binary \cite{2014ApJ...794..167S}. This will add up to the effect we put forward here.


Apart from not considering more involved and detailed accretion models, in our analysis we also did not take into account dynamical friction from the stellar component affecting the very soft binaries neither the many and complex dynamical effects arising from close encounters.  Refinements and more detailed analysis will follow. 
Our main message is that the accretion in primordial clusters seems to have drastic consequences regarding the BH mass function and BBH merger rates in GCs. Not only can it not be ignored, but also we feel urged to consider the mechanism in much more involved and realistic scenarios to quantify more precisely its effect. 

Finally, the possibility that the gas is depleted in proto-globular clusters primarily due to the accretion by dark remnants which leads to the formation of either a dense, highly populated, BH subcluster \cite{Breen_Heggie_2013MNRAS.436..584B,Morscher_2015ApJ...800....9M,Chatterjee_2017ApJ...834...68C,Weatherford_2017arXiv171203979W,Askar-et-al_2018arXiv180205284A} that may form a disk \cite{Roupas_2017ApJ...842...90R,Meiron_2018arXiv180607894M} or an IMBH (for observational evidence see e.g., \cite{Gebhardt_2002ApJ,Gebhardt_2005ApJ,Noyola_2008ApJ,Sun_2013ApJ,Feldmeier_2013AA,
Perera_2017MNRAS,Mezcua_2017arXiv170509667M,2017Natur.542..203K}) needs further investigation. Certainly, our calculations provide indirect support to these scenarios.

\appendix

\section{Accretion of gas by BHs}\label{sec:acc}

The spherically symmetric accretion of gas with density $\rho_g$ by a BH with mass $m_\bullet$ is described by the Bondi formula \cite{Bondi_1952MNRAS.112..195B}, which is an application of the continuity equation
\begin{equation}
        \dot{m}_\bullet = 4\pi r_{inf}^2 \rho_g v
\end{equation}
where $v$ is the inward velocity of the gas and $r_{inf}$ is the radius of influence of the BH, that is, the space within which the BH's gravitational potential dominates the gas dynamics. Assuming
\begin{equation}
        r_{inf} \sim \frac{Gm_\bullet}{v^2}
\end{equation}
we get
\begin{equation}\label{eq:dmdt}
        \dot{m}_\bullet = \frac{4\pi G^2}{v^3} \rho_g m_\bullet^2.
\end{equation} 
We scale Eq. (\ref{eq:dmdt}) as
\begin{equation}\label{eq:dmdt_res}
        \dot{m}_\bullet = 4.75  \left( \frac{5 km/s}{v}\right)^3 \left(\frac{\rho_g}{10^5 M_\odot/pc^3}\right) \left( \frac{m_\bullet}{ 5 M_\odot}\right)^2 \frac{M_\odot}{Myr}.
\end{equation}

Now, for $\rho_g = const$ in time, we get
\begin{equation}\label{eq:m_acc}
        m_\bullet = m_{\bullet,0} \left(1 - \frac{t}{\tau}\right)^{-1}.
\end{equation}
where $m_{\bullet,0} = m_{\bullet}(t=0)$ and
\begin{align}\label{eq:tau}
        \tau &= \frac{m_\bullet}{\dot{m}_\bullet} = \frac{v^3}{4\pi G^2 m_{\bullet,0} \rho_{g}} \\
                &= 1.05 \left(\frac{v}{5km/s}\right)^3 \left( \frac{5 M_\odot}{m_{\bullet,0}}\right) \left(\frac{10^5M_\odot/pc^3}{\rho_{g}}\right) Myr.
\end{align}
This defines the characteristic timescale of the process, when there appears a runaway effect if there is no explicit gas loss and the implicit loss due to accretion is neglected. 

Eddington-limited accretion rate is given by the formula \cite{Rybicki+Lightman_1979rpa..book.....R}
\begin{equation}\label{eq:Eddington}
        \dot{m}_\text{Edd} = \frac{4\pi G }{\kappa \eta c}m,    
\end{equation}
where $c$ is the speed of light, $\kappa$ the electron scattering opacity and $\eta$ the radiation efficiency we assume to be $\eta = 0.1$. This value corresponds to the formation of a gas accretion disk \cite{Shapiro+Teukolski_1983}. We get 
\begin{equation}
        m_\text{Edd} = m(0) e^{t/\tau_\text{Edd}},\;\tau_\text{Edd} = 45Myr.
\end{equation}

\section{Hardening of BBH due to isotropic accretion}\label{sec:hard}

Hadjidemetriou \cite{Hadjidemetriou_1963Icar....2..440H,Hadjidemetriou_1966Icar....5...34H} has shown that isotropic mass increase or loss of a binary can be described by a dragging force
\begin{equation}
        \vec{F}_{acc} = -\frac{1}{2} \dot{m}_b \vec{v}_b 
\end{equation}
acting on the total mass $m_b = m_1 + m_2$, $m_{i=1,2}$ denotes the mass of each member. He calculated, using the Lagrange method of variation of Keplerian elements, that it is in this case
\begin{align}   
        \frac{da}{dt} &= -a \frac{1+2e\cos f +e^2}{1-e^2} \frac{\dot{m}_b}{m_b} \\
        \frac{de}{dt} &= -(e + \cos f) \frac{\dot{m}_b}{m_b} \\
\label{eq:df_dt}
        \frac{df}{dt} &= \frac{2\pi}{P} \frac{(1+e \cos f)^2}{(1-e^2)^{3/2}} 
                        + \frac{\sin f }{e} \frac{\dot{m}_b}{m_b}
\end{align}
where $a$, $e,$ and $f$ are the semi-major axis, the eccentricity and the true anomaly respectively and we denote
\begin{equation}
        P = 2\pi \frac{a^{3/2}}{\sqrt{Gm_b}}
\end{equation}
the period of the orbit. It is straightforward to verify that the specific angular momentum is conserved
\begin{equation}\label{eq:spec_l}
        l = \sqrt{Gm_b a(1-e^2)} = const.
\end{equation}
We then consider the secular evolution of the mean value of any quantity $A(t)$ over the time $P$
\begin{equation}
\left\langle A(t) \right\rangle \equiv \frac{1}{P}\int_0^P A(t) dt
\end{equation}
assuming slow accretion within one period
\begin{equation}\label{eq:slow_acc}
        \frac{dm_b}{dt} \ll \frac{m_b}{P}.
\end{equation} 
Equation (\ref{eq:df_dt}) becomes in this limit 
\begin{equation}
        \frac{1}{P} dt =  \frac{1}{2\pi}\frac{(1-e^2)^{3/2}}{(1+e \cos f)^2} df
\end{equation}
and therefore we get for the mean value of any quantity $A(t)$ within time $P$ that
\begin{equation}
        \left\langle A \right\rangle = \frac{(1-e^2)^{3/2}}{2\pi}\int_0^{2\pi} A \frac{1}{(1+e\cos f)^2} df.
\end{equation}
Thus, we get for the secular evolution of the orbit
\begin{align}
        \left\langle \frac{d a}{dt}\right\rangle &= -\frac{a (1-e^2)^{1/2}}{2\pi}\frac{\dot{m}_b}{m_b}
        \int_0^{2\pi} \frac{1+2e\cos f + e^2}{(1+e\cos f)^2} df
\\
        \left\langle \frac{d e}{dt}\right\rangle &= -\frac{(1-e^2)^{3/2}}{2\pi}\frac{\dot{m}_b}{m_b}
        \int_0^{2\pi} \frac{e + \cos f}{(1+e\cos f)^2} df.
\end{align}
After performing the integration we get
\begin{align}\label{eq:a_Had_sec}
        \left\langle \frac{d a}{dt}\right\rangle &= - a \frac{\dot{m}_b}{m_b}
\\
        \left\langle \frac{d e}{dt}\right\rangle &= 0.
\end{align}
We infer that the eccentricity is secularly conserved and the separation $a$ evolves secularly like
\begin{equation}\label{eq:a_accr}
        a(t) = a(0)\frac{m_{b}(0)}{m_b(t)}.
\end{equation}
The same result we deduce from the conservation of angular momentum (\ref{eq:spec_l}) assuming $e=const.$

However, the crucial quantity, which determines the dynamical evolution of the binary due to close encounters, is not $a$. According to the ``Heggie-Hills law'' \cite{Heggie_1975MNRAS.173..729H,Hills_1975AJ.....80..809H} soft binaries get softer on average and hard ones get harder due to close, mainly thee-body, encounters, in which a hard binary is one with
\begin{equation}
        |E_b| \gtrsim \left\langle \frac{1}{2} m v^2\right\rangle_{cluster}.
\end{equation}
$E_b$ is the internal energy of the binary
\begin{equation}
        E_b = -G\frac{m_1 m_2}{2 a}
\end{equation}
and 
\begin{equation}
        \left\langle E_{kin}\right\rangle_{cluster} = \left\langle \frac{1}{2} m v^2\right\rangle_{cluster}
\end{equation}
 the mean kinetic energy per star of the cluster. Therefore the quantity which determines the dynamical evolution of a binary inside a cluster is the quantity
\begin{equation}\label{eq:chi}
        \chi \equiv \frac{|E_b|}{\left\langle E_{kin}\right\rangle_{cluster}} = \left(\frac{a}{a_h}\right)^{-1},
\end{equation}
we shall call hardness. The quantity $a_h$ is a characteristic separation
\begin{equation}\label{eq:a_h}
        a_h = \frac{G m_1 m_2}{\left\langle m\right\rangle_{cluster} \sigma^2}.
\end{equation}
which follows from equation (\ref{eq:chi}). In the followings we will assume for the mean stellar mass of the primordial cluster $\left\langle m\right\rangle_{cluster} = 0.36 M_\odot$, as estimated recently by Maschberger \cite{Maschberger_2013MNRAS.429.1725M}. The quantity $a_h$ defines the soft-hard boundary $\chi = 1 \Leftrightarrow a=a_h$. The higher the value of $\chi$ the harder the binary. Binaries that have become hard enough, that is, their binding energy is higher than the mean stellar energy $\chi > 1 $, will most probably continue to get harder and eventually merge. Therefore we are interested in determining how soft may a BBH be in order to become hard by accretion before the gas is completely depleted from the cluster.

Using Equations (\ref{eq:a_accr}), (\ref{eq:chi}) we conclude that the binary's hardness increases proportional to the third power of mass due to isotropic accretion
\begin{equation}\label{eq:chi_law}
        \frac{\chi (t)}{\chi (0)} = \left(\frac{m_{b}(t)}{m_{b}(0)}\right)^{3}
        \frac{\gamma (t) (1-\gamma (t))}{\gamma (0)(1-\gamma (0))},\quad
        \gamma (t) \equiv \frac{m_1 (t)}{m_b (t)}
\end{equation}
and therefore it is very sensitive to mass increase. 

\section{Evolution of accreting BBH}\label{sec:evolution}

We wish to estimate the degree that a primordial BBH can become more massive and harder by accretion of primordial gas and by the time the gas is completely depleted. This calculation will allow us to conclude on whether it is possible the BBHs of the GW signals GW150914, GW170104 and GW170814, corresponding to initial BBH masses $m_{BBH} \sim (50-65)M_\odot$ to originate from low mass BBHs ($m_{BBH} \sim 15M_\odot$) that have been grown and hardened rapidly by accretion inside a proto-globular clusters' core.

In our analysis we need some estimation on the gas depletion time and the rate of gas depletion in primordial clusters. Calura et al. \cite{Calura_2015ApJ...814L..14C} estimate that within $\sim 14Myr$ the gas is $99\%$ depleted by star formation feedback processes in a primordial cluster with initial total mass $\sim 10^7M_\odot$. They find that within the first $3Myr$ about $40\%$ of the gas is lost, while the gas is completely depleted by $30Myr$. It is evident in their analysis that there is an approximate linear mass loss of gas for a time lapse $\sim 5Myr$ starting after the first $Myr$, in which time interval occurs the most effective and rapid mass loss while until time $\sim 30Myr$ the rest gas mass is slowly depleted. Therefore, for a linear gas density loss within the time $t_f$ when the gas is completely depleted
\begin{equation}\label{eq:rho_g_t}
        \rho_g(t) = \rho_{g0}\left(1 - \frac{t}{t_f}\right),
\end{equation} 
the more realistic and strict values of $t_f$ should be about $\sim 5Myr$ and up to $\sim 10Myr$. 
For an exponential law like
\begin{equation}\label{eq:rho_g_t-exp}
        \rho_g(t) = \rho_{g0} e^{- \frac{t}{t_e}},
\end{equation} 
the value of $t_e$ should be $\sim 3Myr$ in order to have more than $99\%$ depletion at $\sim 14Myr$. 

We next wish to calculate the minimum initial gas density $\rho_{g0,min}$ inside which a BH of initial mass $m_\bullet(0)$ -member of a BBH- should move, in order to grow in mass by a certain amount within a definite time lapse. Then we can compare with the estimated gas density in primordial clusters $\rho_g \sim 5\cdot (10^4-10^5)M_\odot/pc^3$ (see section \ref{sec:acc}).

We assume that the relative vertical speed $v$ with which the gas is absorbed by the $i$th BBH member equals
\begin{equation}
        v_i(t) = \sqrt{c_s^2 + v_c^2 + v_{\bullet,i}(t)^2},\quad i=1,2
\end{equation} 
where $c_s$, $v_c$,  and $v_{\bullet,i}$ are the speed of sound of the gas cloud, the circular speed of the center of mass of the binary with respect to the cluster center, and the orbital velocity of the member BH of the binary with respect
to the center of mass. This equation follows if one assumes that all directions are equivalent. We assume $c_s = 5km/s$ as estimated previously in section \ref{sec:acc}. For an initially hard  binary $\chi (0) \gtrsim 1$ we assume $v_c = \sigma$ with the stellar velocity dispersion in the core equal to $\sigma = 6km/s$ \cite{B&T_2008gady.book}. The value of $v_{\bullet,i}(0)$ follows from the value of the initial hardness as follows. 
The orbital velocity of the fiducial particle is
\begin{equation}
        v_b = \sqrt{G\frac{m_b}{a}} = \sqrt{G\frac{\chi m_b}{ a_h}}
\end{equation}
where $a_h$ is given in equation (\ref{eq:a_h}). The orbital velocity of each member BH with respect
to the center of mass of the binary is
\begin{equation}
\label{eq:v_member}
        v_{\bullet,1} = \frac{m_2}{m_b} v_b ,\quad
        v_{\bullet,2} = \frac{m_1}{m_b} v_b.
\end{equation}

Assuming $m_1 = m_2 \equiv m_\bullet = m_b/2$ and using equation (\ref{eq:rho_g_t}) we calculate $\rho_{g0,min}$ so that the BH mass is grown $k$ times, meaning
\begin{equation}
        m_\bullet(t_f) > k m_{\bullet}(0) \text{ for any } \rho_{g0} > \rho_{g0,min} 
.\end{equation}
By integrating Equation (\ref{eq:dmdt}), we get for the linear gas loss case (\ref{eq:rho_g_t})
\begin{align}\label{eq:rho_min}
        \rho_{g0,min}^\text{lin} = \frac{(c_s^2 + v_c^2)^{3/2}}{2\pi G^2 m_{\bullet,0} t_f}
        &\left\lbrace 
        \textstyle
        _2F_1\left( (-\frac{3}{2},-\frac{1}{2}),\frac{1}{2},-\frac{v_{\bullet}(0)^2}{c_s^2+v_c^2} \right)
         \right.
         \\
         &- \displaystyle\frac{1}{k} 
\left.   
\textstyle 
        _2F_1\left( (-\frac{3}{2},-\frac{1}{2}),\frac{1}{2},-k^2\frac{v_{\bullet}(0)^2}{c_s^2+v_c^2} \right)         \right\rbrace
\end{align}
where $_2F_1$ is the hypergeometric function. 
For the exponential gas loss (\ref{eq:rho_g_t-exp}) we get
\begin{equation}\label{eq:rho_min-exp}
        \rho_{g0,min}^\text{exp} = \frac{t_f}{t_{A}} \frac{\ln(1-A/100)^{-1}}{2A/100}         \rho_{g0,min}^\text{lin},
\end{equation}
where $t_{A}$ is the time by which $A\%$ of the gas density is depleted. Therefore, for the same initial gas density we get
\begin{equation}
        t_{99} = 2.3 t_f,
\end{equation}
where $t_{99}$ denotes the time by which $99\%$ of the gas density is depleted. 

The evolution of each BBH member may be calculated by solving the system of equations
\begin{align}
        \label{eq:dm1dt}
        \dot{m}_1 (t) = \frac{4\pi G^2 \rho_g(t)}{v_{\bullet,1}^{3}} m_1(t)^2 \\
        \label{eq:dm2dt}
        \dot{m}_2 (t) = \frac{4\pi G^2 \rho_g(t)}{v_{\bullet,2}^{3}} m_2(t)^2 
\end{align}
where
\begin{align}
        v_{\bullet,1} (t) = \sqrt{ c_s^2 + v_c^2 + \frac{m_2(t)^2}{m_b(0)^2}v_b(0)^2 } \\
        v_{\bullet,2} (t) = \sqrt{ c_s^2 + v_c^2 + \frac{m_1(t)^2}{m_b(0)^2}v_b(0)^2 }
\end{align}
and $a_h(t)$ is given in equation (\ref{eq:a_h}). 

\section{Ionization vs accretion}\label{sec:ion}

The ionization (desolution by a single encounter) timescale of a soft BBH is \cite{Heggie_1975MNRAS.173..729H}
\begin{equation}\label{eq:ion}
        t_\text{ion} = \frac{3}{20\sqrt{2\pi}}\frac{m_\bullet}{\left\langle m\right\rangle_{cluster}}\frac{\sigma}{G\rho_\text{stars} a} = \chi \cdot 0.085 \frac{\sigma^3}{G^2\rho_\text{stars} m_{\bullet}},
\end{equation}
where we used the definition of $\chi$ and $a_h$, Eqs. (\ref{eq:chi}), (\ref{eq:a_h}) and assumed equal masses for the BBH members and mass $\left\langle m\right\rangle_{cluster}$ for the single scattering star.
The evaporation (desolution by numerous encounters) timescale is of the same order of magnitude 
$t_\text{ev} \propto \sigma/G\rho_\text{stars} a \ln\Lambda$ \cite{B&T_2008gady.book}. 

Combining Equations (\ref{eq:m_acc}), (\ref{eq:tau}), and (\ref{eq:a_accr}) we find the timescale $\tau_h$ at which a soft binary ($\chi < 1$) may get hard ($\chi \geq 1$) due to Bondi accretion
\begin{equation}\label{eq:tau_h}
        \tau_h = (1-\chi)\cdot 0.08 \frac{v^3}{G^2 \rho_{gas} m_{\bullet}}.
\end{equation}
Dividing (\ref{eq:ion}) by (\ref{eq:tau_h}) we get
\begin{equation}\label{eq:tau_ratio}
        \frac{t_\text{ion}}{\tau_h} = \frac{\chi}{1-\chi} \frac{\sigma^3}{v^3} \frac{\rho_{gas}}{\rho_\text{stars}}.
\end{equation}

\bibliography{2018_Gas_accretion_BBH_GCs}
\bibliographystyle{myunsrt}

\end{document}